\documentstyle[preprint,revtex]{aps}
\begin{document}
\draft
\preprint{}
\begin{title}
Exact phase diagrams for an Ising model on a two-layer Bethe lattice
\end{title}

\author{Chin-Kun Hu \cite{huck},
 N. Sh. Izmailian \cite{nick}}
\begin{instit}
Institute of Physics, Academia Sinica, Nankang, Taipei 11529, Taiwan
\end{instit}

\author{K. B. Oganesyan}
\begin{instit}
Yerevan Physics Institute, Yerevan 375036, Armenia
\end{instit}
\date{\today}

\begin{abstract}
Using an iteration technique, we obtain exact expressions for
the free energy and the magnetization of an Ising model on a two - layer
Bethe lattice with intralayer coupling constants $J_1$ and $J_2$ for the
first and the second layer, respectively, and interlayer coupling
constant $J_3$ between the two layers;
the Ising spins also couple with external magnetic fields, which are
different in the two layers.  We obtain exact phase diagrams for the
system and find that when $|J_3| \to 0$,
$ \Delta T_c \equiv \frac{T_c(J_3)-T_c(0)}{T_c(0)} \sim
|J_3/J_1|^{1/{\psi}}$, where $T_c(J_3)$ is the phase
transition temperature for the system with interlayer
coupling constant $J_3$ and the shift exponent $\psi$ is 1
for $J_1 = J_2$ and is 0.5 for $J_1 \ne J_2$. Such results
are consistent with predictions of a scaling theory.
We also derive equations for $ \Delta T_c $ when $|J_3|$ approaches
$\infty$.  
\end{abstract}

\newpage
\narrowtext

\begin{center}
I. INTRODUCTION
\end{center}

The physical properties of various magnetic layered structures and
superlattices have been intensely studied both experimentally and
theoretically for reasons ranging from fundamental investigations of phase
transitions to technical problems encountered in thin-film magnets
\cite{r1}. Experimentally, sub-monolayer and monolayer films of ferromagnetic
materials offer challenging opportunities to fabricate materials with various
novel magnetic properties, such as giant magnetoresistance, surface magnetic
anisotropy, enhanced surface magnetic moment and surface magneto-elastic
coupling. On theoretical grounds, surface
magnetism has been treated within several different frameworks: mean-field
approximations \cite{r2}, effective-field theories \cite{r3}, spin-fluctuation
theory \cite{r5}, renormalization-group methods \cite{r6}, two-site cluster
approximations \cite{r7} and Monte-Carlo techniques \cite{r8}. Though each
method has it own advantages, there all have limitations in 
treating film systems. Numerical techniques such as Monte Carlo 
method can provide very accurate results for properties of finite systems; 
however, they are computation-intensive and can be carried out only for 
relatively small system sizes.

Since exact solutions for realistic layer systems on regular lattices are 
generally unavailable, one relies on approximation 
schemes to obtain a qualitative picture of the phase diagram. 
In our approach, we replace the original two-layer regular lattice by a 
two-layer Bethe lattice with the same coordination number $q$ as the original 
lattice. Once this approximation is made, we can solve our model exactly
as in the case of one-layer Bethe lattice \cite{baxter82}. 
It is now widely recognized (see e.g. \cite{kt91,m92,m94,r9}) that in 
many cases solutions of a spin model on Bethe or generalized Bethe 
lattices are qualitatively better approximations for
the regular lattice than solutions obtained by 
conventional mean-field theories, because of the presence of correlations
(albeit weak ones) in the former \cite{ih98} 
and the lack of correlations in the latter. 
It has also been found that phase diagrams of an Ising model on
a Husimi tree (a Bethe-like lattice) with two-spin and three-spin 
interactions \cite{m92,m94} closely match exact phase diagrams of an 
Ising model on a two-dimensional Kagome lattice with two-spin and three-spin 
interactions \cite{ww89}.
Of course, our approximation also has limitations:   
since correlations are weak \cite{ih98}, it predicts a
transition temperature which is higher than that for a regular lattice,
and it is usually not reliable for predicting critical exponents. 
On the other hand, the Monte-Carlo method will be highly reliable
for predicting critical exponents. But we believe that  
our approach correctly gives the general shape of the phase diagram.

It is well known that
the Bethe and Bethe-like lattices cannot be embedded in a finite-dimensional 
Euclidean space without distortion in their bond angles and lengths 
\cite{hs82}. On the other hand, it has been pointed out by Mosseri and Sadoc 
\cite{ms82} that these structures can be considered as regular lattices 
of fixed bond angles and lengths 
if they are embedded in a two-dimensional space of 
constant negative curvature (the hyperbolic or Lobachevsky plane ${\bf H2}$ 
\cite{c47}). The 
surface of negative curvature are now being introduced to describe some 
complex structure with large cells, formed by inorganic and organic 
materials, which can be considered as crystals of surface and films. Among 
them are the cubic crystalline structures formed by amphiphilic molecules in 
the presence of water \cite{sc89,sc90} and a magnetically coupled 
three-dimensional (Terephthalato)  manganese(II) network \cite{hd97,k93}. 
Furthermore, non-Euclidean hyperbolic symmetries have even been found in 
hexagonal and cubic-close-packed Euclidean crystals \cite{j91}. The local 
structural similarities between that materials and negatively curved Bethe 
lattice suggest that the two-layer Bethe lattice considered in the present
paper can be related to some physical systems. 

In this paper, we use an iteration technique to obtain exact expressions for
the free energy and the magnetization of an Ising model on a two - layer
Bethe lattice
with intralayer coupling constants $J_1$ and $J_2$ for the
first and the second layer, respectively, and interlayer coupling
constant $J_3$ between two layers;
the Ising spins also couple with external magnetic fields, which are
different in the two layers.
Allowing for a difference
in these two fields is important, because they act in an
opposing manner on the zero-field boundaries \cite{r13}.
We obtain exact phase diagrams for the system and
find that when $|J_3| \to 0$,
$ \Delta T_c \equiv \frac{T_c(J_3)-T_c(0)}{T_c(0)} \sim
|J_3/J_1|^{1/{\psi}}$, where $T_c(J_3)$ is the phase
transition temperature for the system with interlayer coupling
constant $J_3$ and the shift exponent $\psi$ is 1
for $J_1 = J_2$ and is 0.5 for $J_1 \ne J_2$. Such results
are consistent with predictions of a scaling
theory \cite{abe70,suzuki71,ht97}.
We also derive equations for $ \Delta T_c $ when $|J_3|$ approach
$\infty$.  

The theoretical works on thin-layer systems  before the present paper
were less general. The system of coupled two-dimensional Ising planes
on regular lattices, e.g. the square lattice,
is not exactly soluble; however, it has been investigated by a variety
of approximate methods.  Ballentine \cite{ballentine64}
used high-temperature series expansions to study 
the model with $J_1=J_2=J_3$ \cite{J123}.  This work was later
extended by Allan \cite{allan70} to films up to five layers and by Capehart
and Fisher \cite{cf76} to films up to ten layers.
The two - layer system where the inter-layer
coupling constant differs from the intra-layer coupling constant
was studied by Abe \cite{abe70} in the context of a scaling theory
valid in the limit of a weak interlayer coupling.
The more general case in which
$J_1 \ne J_2$ has also received some attention. The most complete treatment
was that of Oitmaa and Enting \cite{oe75}, who combined mean-field theory,
scaling theory and high-temperature expansions in a study of the two-layer
model, and calculated the variation of the critical temperature, the
layer magnetizations and the interlayer correlation function with $J_3$.
Recently, Ferrenberg and Landau \cite{fl91} considered the same two-layer
problem using Monte Carlo simulations and mean-field theory.
The Bethe lattice version of a thin film was first studied in \cite{r12},
in which the phase diagrams of coupled bilayers with $J_1 = J_2$ and
zero external fields were obtained.

More recently, there have been many investigations of the two - layer
Ising model due to stimulation from experiments
for ultra - thin magnetic films
\cite{ht97,ls93,hlim93,lsu93,w94,bw94,hlt96,acpv92,us94,mu96,bg97}.
The experimental study of such systems has made significant advances in recent
years due to improvements in surface-force apparatus and microscopic
techniques.  Many theoretical studies were devoted to the model with two
exchange parameters, i.e. $J_s$ for spins on free surfaces and $J$ for
all other spins. However, in our opinion the exchange interaction $(J_3)$
between the surface and the second layer has an important influence on
the surface magnetic order.
Therefore, in the present work, we study an Ising model
on a two-layer Bethe lattice with three exchange parameters
$J_1$, $J_2$ and $J_3$ and in the presence of magnetic
fields, which are different in the two layers.
We will calculate the free energy of that system and investigate its
critical properties.

In previous papers \cite{r14,aio98}, the relations between
the free energies of a spin-1 Ising model on Bethe and Cayley trees
and of a multi-site Ising model on Husimi lattices and generalized Cayley
trees are obtained. In that approach one obtains the free
energy from recursion relations, and equations for physical
quantities by differentiation of the free energy
functionals with respect to external fields.
Recently, an elegant and original method of computing the
bulk free energy for any model on infinite Bethe and Husimi lattices was
presented by Gujrati \cite{r9}.
In the present paper, we extend this method to study an Ising model
on a bilayer Bethe lattice. In particular, we will calculate
exact phase diagrams and the shift exponent $\psi$ for the system.

In 1970 - 1971, Abe \cite{abe70} and Suzuki \cite{suzuki71} used
 a scaling theory to predict that the shift exponent $\psi$ for the
two-layer planar lattice Ising model is equal to the susceptibility 
exponent ($\gamma = 1.75$) of the two-dimensional Ising model.
In 1992, Angelini et al. predicted that $\psi = 1.5$ \cite{acpv92};
in 1993, Lipowski and Suzuki predicted that $\psi=2$ \cite{lsu93}.
Very recently, Horiguchi and Tsushima
\cite{ht97} used a $\beta$ - function approach \cite{barber83} 
to obtain $\psi = 1.73 \pm 0.04$ for the two - layer square lattice 
Ising model with $J_1=J_2$, which is very close to
the theoretically predicted value $\psi = \gamma = 1.75$.
They also found that the shift exponent $\psi$ for the system with
$J_1 \ne J_2$ is $0.5$ and explained this value
in terms of a scaling theory. In 1998 Lipowski \cite{l98} used a 
transfer-matrix mean-field approximation to calculate
the shift exponent $\psi$. He found $\psi=1.79$ for the system 
with $J_1=J_2$ and $\psi=0.501$ for the system with
$J_1 \ne J_2$. 
 
In this paper, we consider
both $J_1 = J_2$ and $J_1 \ne J_2$ for the Ising model on a two-layer
Bethe lattice and
obtain exact values of $\psi$, which are consistent with the predictions
of the scaling theory.

The outline of this paper is as follows. In Sec. II, we present
the bilayer Ising model and discuss different types of ground states. In
Sec. III, we derive the exact free energy, equations of state and order
parameters for the Ising model on the bilayer Bethe lattice.
In Sec. IV, we investigate temperature dependence of the order
parameters and discuss the phase diagrams. In Sec. V, we calculate the
critical temperature in weak and strong vertical coupling regimes
and obtain exactly the shift exponent $\psi$.
In Sec. VI, we conclude with some general remarks
concerning our results.

\begin{center}
II. TWO-LAYER MODEL AND ITS GROUND STATES
\end{center}

In this section, we consider an ultra-thin film composed  of two
atomic layers $G_1$ and $G_2$ such that the lattice structures
of $G_1$ and $G_2$ are identical and each of them have $N$ sites
and coordination number $q$; the corresponding lattice sites in
$G_1$ and $G_2$ are labeled by
$i$ and $i'$, respectively, where $1 \le i, i' \le N$ and they are
nearest neighbors (nn) to each other.
The Ising Hamiltonian on such two-layer lattice can be written as
\begin{equation}
\label{R1}
-{\beta} H=J_1\sum_{<ij>}s_is_j + J_2\sum_{<i'j'>}{\sigma}_{i'}{\sigma}_{j'} +
J_3\sum_{<ii'>}{s_i\sigma}_{i'} + h_1\sum_is_i + h_2\sum_{i'}{\sigma}_{i'} ,
\label{H}
\end{equation}
where $\beta=(k_B T)^{-1}$ with $k_B$ being the Boltzmann constant and
$T$ being the temperature,  $s_i$ and ${\sigma}_{i'}$
take values $\pm 1$, $J_1$ and $J_2$ are, respectively, the
coupling constants of the exchange interaction between the pair of
nn spins in the first and the second layer,
$J_3$ is the coupling constant between a spin in the first layer
and its nn in the second layer, and $h_1$ and $h_2$
are magnetic fields acting on spins in the first and the
second layer, respectively.

This model has three order parameters, two of which correspond to
the thermal average of total spins of the first and the second
layers, respectively,
\begin{equation}
\label{m1}
m_1=\frac{1}{N}\sum_{i=1}^N<s_i>,
\quad \quad
m_2=\frac{1}{N}\sum_{i'=1}^N<\sigma_{i'}>.
\end{equation}
These order parameters can be defined by variation of the partition
function with respect to $h_1$ and $h_2$. The total magnetization density
$m$ and the density of staggered magnetization $\eta$
are defined by

\begin{equation}
\label{mm}
m ={1 \over 2}(m_1+m_2), \quad \quad \eta ={1 \over 2} (m_1 - m_2).
\end{equation}
The third order parameter corresponds to the interlayer spin-spin
correlation function between nn spins of adjacent layers
\begin{equation}
\label{rho}
\rho =\frac{1}{N}\sum_{i=i'=1}^N(<s_i\sigma_{i'}>-<s_i><\sigma_{i'}>).
\end{equation}

Before studying the temperature dependences of the order parameters,
let us investigate the ground states of the model at $T=0$
analytically.
The ground-state energy in units of $|J_1|$ and in the absence of 
magnetic fields may be described by the following Hamiltonian

\begin{equation}
\label{E}
E = -\sum_{<plaq>}\left[\frac{J_1}{|J_1|}s_is_j +
\frac{J_2}{|J_1|}\sigma_{i'}\sigma_{j'} +
\frac{J_3}{q|J_1|}(s_i\sigma_{i'}+s_j\sigma_{j'})\right].
\end{equation}
Here the summation goes over all plaquettes and each plaquette consists
of four nearest-neighbor pairs of the two-layer system with one pair, 
$<ij>$, on $G_1$, one pair, $<i'j'>$, on $G_2$, and two pairs, 
$<ii'>$ and $<jj'>$, connecting $G_1$ and $G_2$.

By comparing the values of $E$ for different spin configurations, 
we obtain the ground-state phase diagrams shown in Figs. 1(a) 
and 1(b) for $J_1 > 0$ and $J_1 < 0$, respectively.
We find five types of ground states with following values of
the order parameters ($m, \eta, \rho$):

$$
(I) \qquad \qquad \qquad \quad  m = \pm 1, \quad  \eta = 0, \quad \rho = 0, 
$$
$$
(II) \qquad \qquad \qquad \quad m = 0, \quad \eta = \pm 1, \quad \rho = 0, 
$$
$$
(III)\qquad \qquad \qquad \quad m = 0, \quad \eta = 0, \quad \rho = 1, 
$$
$$
(IV)\qquad \qquad \qquad \quad m = 0, \quad \eta = 0, \quad \rho = -1, 
$$
$$
(V)\qquad \qquad \qquad  m = \pm 1/2, \quad \eta = \pm 1/2, \quad \rho = 0. 
$$

The coordinates ($J_2/|J_1|, J_3/q|J_1|$) of the multiphase points are:
\begin{equation}
A_1 \to (0, 0), \quad B_1 \to (-1, 1) \quad  \mbox{and}  \quad
C_1 \to (-1, -1),  \quad \quad \quad \quad \mbox{for} \quad J_1>0
\label{point1}
\end{equation}
and
\begin{equation}
A_2 \to (0, 0),  \quad B_2 \to (1, 1)  \quad \mbox{and}  \quad
C_2 \to (1, -1), \quad \quad \quad \quad \mbox{for} \quad J_1 <0.
\label{point2}
\end{equation}

Phase (I) represents the usual ferromagnetic 
ordering $m_1=m_2$ ($\eta=0$).

Phase (II) represents ferromagnetic ordering in $G_1$ and $G_2$,
but magnetizations in $G_1$ and $G_2$ are antiparallel,
i.e. $m_1=-m_2$ and $m=0$ 
(interlayer ordering is antiferromagnetic type).
It is worthwhile to note that this phase corresponds to the well known
compensation phenomenon which occurs when the
magnetizations of two layers cancel each other instead of being equal.

Phase (III) represents the antiferromagnetic ordering
in both layers ($m_1 = m_2 = 0$)
where interlayer ordering is ferromagnetic ($\rho=1$).

Phase (IV) represents the totally 
antiferromagnetic ordering ($\rho=-1$).

Phase (V) represents the ferromagnetic ordering ($m=\pm 1/2,
\eta=\pm1/2$), which is equivalent to
the case that the ground state of one layer is
ferromagnetic and the ground state of another layer is antiferromagnetic.

The phases (I) - (V) will be referred to as (F) - ferromagnetic, (C) -
compensated, (M) - mixed, (A) - antiferromagnetic and (SF) - surface
ferromagnetic phase, respectively. 

\begin{center}
III. EQUATIONS OF STATE AND FREE ENERGY.
\end{center}

Let us consider an Ising model on a bilayer Bethe lattice,
which is constructed by connecting to the
central pair of sites $q$ pairs in order to form the first
generation and by connecting successively $(q-1)$ pairs to each pair
in a generation to form the next generation.
The result is an infinite lattice in which every site has
$(q+1)$ nearest neighbors, where $q$ nearest neighbors are in the
same layer as the site and one nearest neighbor is in another layer.

The partition function of the system represented by
Eq. (\ref{H}) may be written as

\begin{equation}
\label{R2}
Z=\sum_{\{\sigma,s \}} \exp{\left\{J_1\sum_{<ij>}s_is_j +
J_2\sum_{<i'j'>}{\sigma}_{i'}{\sigma}_{j'} 
+ J_3\sum_{<ii'>}{\sigma}_{i'}s_i 
+ h_1\sum_is_i + h_2\sum_{i'}{\sigma}_{i'}\right\}} ,
\end{equation}
where the sum goes over all configurations of the system.

Now we derive exact recursion relations for $Z$.
When the Bethe tree is cut apart
at the central pair, it separates into $q$ identical branches, each
of which contains $(q-1)$ branches. The partition function can be written
as follows:

\begin{equation}
\label{R3}
Z=\sum_{\{{\sigma}_0, s_0 \}}\exp{\left\{J_3{\sigma}_0 s_0 + h_1 s_0 +
h_2{\sigma}_0\right \}}
g_n^{q}({\sigma}_0,s_0),
\end{equation}
where ${\sigma}_0$ and $s_0$ are the spins of the central pair,
$n$ is the number of generations ($n \to \infty$ corresponds
to the thermodynamic limit where
surface effects may be neglected) and $g_n(\sigma_0,s_0)$ is
the partition function of a separate branch. Each branch, in turn,
can be cut apart at the pair of sites nearest to the central pair. The
expression for $g_n({\sigma}_0,s_0)$ can therefore be written in the
following form

\begin{equation}
\label{R4}
g_n({\sigma}_0,s_0)=\sum_{\{{\sigma}_1,s_1\}} \exp{\left \{
J_1s_0s_1 + J_2{\sigma}_0{\sigma}_1 +
J_3{\sigma}_1s_1 + h_1s_1 + h_2{\sigma}_1\right \}}
g_{n-1}^{q-1}({\sigma}_1,s_1).
\end{equation}

Let us introduce the following variables $x_n$, $y_n$ and $t_n$

$$
x_n=\frac{g_n(++)}{g_n(--)}, \quad
y_n=\frac{g_n(+-)}{g_n(-+)}, \quad
t_n=\frac{g_n(-+)}{g_n(--)}.
$$
>From Eq.(\ref{R4}) we easily obtain the recursion relations:

$$
x_n = f_1(x_{n-1},y_{n-1},t_{n-1}), 
$$
\begin{equation}
y_n = f_2(x_{n-1},y_{n-1},t_{n-1}), \label{recrl1} 
\end{equation}
$$
t_n = f_3(x_{n-1},y_{n-1},t_{n-1}), 
$$
\noindent{}
where

$$
f_1(x_n,y_n,t_n)=\frac{A_n\exp{(J_1+J_2)}+\exp{(-J_1-J_2)}
 +D_n\exp{(J_1-J_2)}+B_n\exp{(-J_1+J_2)}}
{A_n\exp{(-J_1-J_2)} + \exp{(J_1+J_2)} + D_n\exp{(-J_1+J_2)} +
B_n\exp{(J_1-J_2)}},
$$
$$
f_2(x_n,y_n,t_n)=\frac{A_n\exp{(-J_1+J_2)} + \exp{(J_1-J_2)} 
 + D_n\exp{(-J_1-J_2)} + B_n\exp{(J_1+J_2)}} {A_n\exp{(J_1-J_2)}
 +\exp{(-J_1+J_2)}+D_n\exp{(J_1+J_2)}+ B_n\exp{(-J_1-J_2)}},
$$
$$
f_3(x_n,y_n,t_n)=\frac
{A_n\exp{(J_1-J_2)}+\exp{(-J_1+J_2)}+D_n\exp{(J_1+J_2)}+
B_n\exp{(-J_1-J_2)}}
{A_n\exp{(-J_1-J_2)} + \exp{(J_1+J_2)} + D_n\exp{(-J_1+J_2)}
+ B_n\exp{(J_1-J_2)}},
$$
\noindent{} 
with
$$
A_n=x_n^{q-1}\exp{(2h_1+2h_2)}, \quad D_n=t_n^{q-1}\exp{(-2J_3+2h_1)}, 
 \quad B_n=y_n^{q-1}t_n^{q-1}\exp{(-2J_3+2h_2)}.
$$

Through $x_n$, $y_n$, and $t_n$ one can express the magnetization and other
thermodynamic quantities, so we can say that in the thermodynamic limit 
($n \to \infty$) $x_n$, $y_n$ and $t_n$ determine the states of the system.
For this reason the recursion relations can also be called the equations of
state (EOS) for the two-layer Ising model. The
magnetizations of the first and the second layers as well
as the spin-spin correlation function
between spins of adjacent layers are expressed by:

\begin{equation}
m_1 = <s_0>= \frac{x_nA_n-1+t_nD_n-y_nt_nB_n}
 {x_nA_n+1+t_nD_n+y_nt_nB_n}, \label{ord1} 
\end{equation}

\begin{equation}
m_2 = <\sigma_0> = \frac{x_nA_n-1-t_nD_n+y_nt_nB_n}
 {x_nA_n+1+t_nD_n+y_nt_nB_n},  \label{ord2} 
\end{equation}

\begin{equation}
<\sigma_0s_0> = \frac{x_nA_n+1-t_nD_n-y_nt_nB_n}
 {x_nA_n+1+t_nD_n+y_nt_nB_n}. \label{ord3}
\end{equation}

We are interested in the case when $(x_n,y_n,t_n)$ converges to a stable
point $(x_s,y_s,t_s)$, which is associated with the thermodynamic solutions
of the two - layer Ising model. In this case the recursion
relations (or equations of state) given by Eq. (\ref{recrl1}) can be
rewritten in the following form:

\begin{equation}
\left(\frac{1-y}{1+y}\right)^{q-1}\exp{(2h_2-2h_1)}=
\frac{u_1-v_1}{u_1+v_1}, \label{ey}
\end{equation}

\begin{equation}
\left(\frac{1-t}{1+t}\right)^{q-1}\exp{(2h_1+2h_2)}=
\frac{u_2-v_2}{u_2+v_2},\label{et}
\end{equation}

\begin{equation}
x^{2(q-1)}\left(\frac{1-y^2}{1-t^2}\right)^{q-1}\exp{(-4J_3)}=
\frac{u_1^2-v_1^2}{u_2^2-v_2^2},\label{exyt}
\end{equation}
where
$$
u_1=c_1x-c_2, \quad v_1=s_1xy+s_2t, \quad c_1=\cosh{(J_1+J_2)},  \quad
s_1=\sinh{(J_1+J_2)},
$$
$$
u_2=c_1-c_2x, \quad v_2=s_1t+s_2xy, \quad
c_2=\cosh{(J_1-J_2)}, \quad s_2=\sinh{(J_1-J_2)},
$$
and

$$
x=\frac{1+y_s}{1+x_s}t_s; \quad y=\frac{1-y_s}{1+y_s};
\quad t=\frac{1-x_s}{1+x_s}.
$$

The total magnetization density ($m=(m_1 + m_2)/2$), 
the density of the staggered
magnetization ($\eta=(m_1 - m_2)/2$) and the density of 
the interlayer spin-spin
correlation function ($\rho = <{\sigma}_0s_0> - m_1m_2$) can be expressed as

\begin{equation}
m = -\frac{tu_2+v_2}{u_2+tv_2+xu_1+v_1xy}, \label{mm12} 
\end{equation}
\begin{equation}
\eta = \frac{(xyu_1+xv_1)}{u_2+tv_2+xu_1+v_1xy}, \label{eta} 
\end{equation}
\begin{equation}
\rho = \frac{(1-t^2)(u_2^2-v_2^2) - x^2(1-y^2)(u_1^2-v_1^2)}
{(u_2+tv_2+xu_1+xyv_1)^2}. \label{rho1}
\end{equation}
In the case when $(x_n,y_n,t_n)$ converge to a stable point
$(x_s,y_s,t_s)$, we can obtain
an equation for the free energy functional $F$:

$$
-\beta F = -\frac{1}{8}\ln{(u_1^2-v_1^2)(u_2^2-v_2^2)} +
 \frac{q-1}{8}\ln{x^2(1-t^2)(1-y^2)} 
$$
\begin{equation}
-\frac{q-2}{4}\ln{(u_2+tv_2+xu_1+xyv_1)} + \frac{1}{2}\ln{2}
+\frac{q}{4}\ln{|c_1^2-c_2^2|}.
\label{free3}
\end{equation}
In deriving this equation we have used the exact relation between the free
energy of the Bethe lattice and Cayley trees \cite{r9,r14,aio98}.

It is easily seen that the expressions for the order parameters $m_1$, $m_2$
and $\rho$ can be obtained by differentiation of the free energy functional
of Eq. (\ref{free3}) with respect to the magnetic fields $h_1$, $h_2$
and the coupling constant $J_3$, respectively. In this sense, 
the interlayer coupling constant, $J_3$,
is analogous to an external field.

This result for the free energy is very useful for locating phase
transitions in case of multiple solutions of the equation of state and for
determining the equilibrium state. Using this free energy functional one can
obtain the full phase diagram, describing not only the  
continuous phase transitions but also the discontinuous ones.

In the next section we will discuss the critical properties of our
model; in particular, we will calculate the critical temperature as a
function of ratios of coupling constants and will show the full phase
diagram in the three-dimensional parameter space spanned by coupling
constants $J_1, J_2$ and $J_3$, for different values of the
coordination number $q$.

\begin{center}
IV. PHASE DIAGRAMS
\end{center}

Now we consider the critical properties of the Ising model
on a two-layer Bethe lattice with different ferromagnetic coupling
constants ($J_1 > 0, J_2 > 0$). Without loss of generality, we need only
consider $J_1 \ge J_2$. The phase transition occurs when $h_1=h_2=0$. In this
case Eqs. (\ref{ey}), (\ref{et}) and (\ref{exyt}) become:

\begin{equation}
x^{2(q-1)}\left(\frac{1-y^2}{1-t^2}\right)^{q-1}\exp{(-4J_3)}=
\frac{u_1^2-v_1^2}{u_2^2-v_2^2},\label{exyt0}
\end{equation}

\begin{equation}
(1+y)^{q-1}(u_1 - v_1) = (1-y)^{q-1}(u_1 + v_1) \quad
\Longleftrightarrow \quad v_1 = yu_1{\Phi}(y^2),
\label{ey0}
\end{equation}

\begin{equation}
(1+t)^{q-1}(u_2 - v_2) = (1-t)^{q-1}(u_2 + v_2) \quad
\Longleftrightarrow \quad v_2 = tu_2{\Phi}(t^2),
\label{et0}
\end{equation}
where

\begin{equation}
{\Phi}(x) = \left(\sum_{n=0}^{[\frac{q-2}{2}]}{C_{2n+1}^{q-1}x^{n}}\right)
/ \left(\sum_{n=0}^{[\frac{q-1}{2}]}{C_{2n}^{q-1}x^{n}}\right), \quad
\quad  \quad  \quad  {\Phi}(0) = q-1,
\label{ey2}
\end{equation}
and $C_{k}^{q-1}$ is a binomial coefficient.

It can be seen that there always exists one solution
of this system:

$$
(i) \quad \quad \quad \quad y = t = 0, \quad \mbox{and}
\quad x^{q-1}\exp{(-2J_3)}=\frac{u_1}{u_2}.
$$
This solution corresponds to the high temperature 
paramagnetic phase ($m_1=0$, $m_2=0$).
In addition to the first solution, the equations of state also have other
solutions with $y\ne0, \quad t\ne0$:
$$
 \quad \quad \quad \quad
s_2^2x = \left[u_1{\Phi}(y^2) - s_1 x\right]
\left[u_2{\Phi}(t^2) -s_1\right], 
$$
$$
(ii) \quad \quad \quad \quad
t^2\left[u_2{\Phi}(t^2) - s_1\right]=xy^2
 \left[u_1{\Phi}(y^2) - s_1 x\right], 
$$
$$
\quad \quad \quad \quad
x^{2(q-1)}\left(\frac{1-y^2}{1-t^2}\right)^{q-1}\exp{(-4J_3)}=
\frac{u_1^2-v_1^2}{u_2^2-v_2^2}.
$$
Of course, only the solution which minimizes the free
energy functional (\ref{free3}) is thermodynamically stable.
The others correspond to unstable or metastable states. If there are two or
more solutions which have the same minimum free energies, these
phases coexist and the system has a first-order phase transition.

When these two solutions merge into one solution,
i.e. $(i)$ = $(ii)$, we obtain the critical line of second-order phase
transitions
\begin{equation}
\label{line}
\exp{(2J_3)}=\frac{c_1-c_2x_{\lambda}}{c_1x_{\lambda}-c_2}x_{\lambda}^{q-1}.
\end{equation}
where $x_{\lambda}$ is the solution of the following equation
\begin{equation}
\label{xlambda}
s_2^2x_{\lambda}=[((q-1)c_1-s_1)x_{\lambda}-(q-1)c_2]
[(q-1)c_1-s_1-(q-1)c_2x_{\lambda}].
\end{equation}
It is convenient now to introduce the new parameters $k_1$ and $k_2$
$$
k_1 = \sqrt{\tanh{J_1}\tanh{J_2}}, 
$$
\begin{equation}
k_2 = \sqrt{\tanh{(-J_1+0.5\ln{\frac{q}
{q-2}})}\tanh{(-J_2+0.5\ln{\frac{q}{q-2}})}}. \label{k1k2}
\end{equation}
The two solutions $x_{\lambda}^{(1,2)}$
of the Eq. (\ref{xlambda}) can thus be
expressed as
\begin{equation}
\label{x1x2}
x_{\lambda}^{(1)}=\frac{1-k_1k_2}{1+k_1k_2},
\quad x_{\lambda}^{(2)}=\frac{1+k_1k_2}{1-k_1k_2},
\end{equation}
and the corresponding expressions for the
$\lambda$-lines of the second-order
phase transition in  the three-dimensional parameter space spanned by
$J_1, J_2$ and $J_3$ will take the form
\begin{equation}
\label{line1}
\exp{\left(2J_3^{(1)}\right)}=
\frac{k_1+k_2}{k_1-k_2}\left(\frac{1-k_1k_2}{1+k_1k_2}\right)^{q-1}, \quad
\quad J_3 > 0 ;
\end{equation}

\begin{equation}
\label{line2}
\exp{\left(2J_3^{(2)}\right)}=
\frac{k_1-k_2}{k_1+k_2}\left(\frac{1+k_1k_2}{1-k_1k_2}\right)^{q-1},\quad
\quad J_3 < 0. 
\end{equation}
The critical lines of the second-order phase transition given by Eqs.
(\ref{line1}) and (\ref{line2}) separate the paramagnetic (P) phase from
the ferromagnetic (F) and compensated (C) phases, which are in turn
separated by a first-order phase transition line.

Before discussing the phase diagram, it is convenient to introduce the
parameters $n = J_2/J_1$, $\delta = J_3/qJ_1$ and $T=J_1^{-1}$.
In terms of the $T, n$ and $\delta$, Eqs. (\ref{line1}) and
(\ref{line2}) for the  $\lambda$ - lines imply a relation
$T = T_c(n, \delta)$ which locates the  critical temperature  as a
function of $n$ and $\delta$ for arbitrary values of the coordination
number $q$. The two critical lines start at
\begin{equation}
\label{Tmax}
T_c^{max} = \frac{2(1+n)}{\ln{[q/(q-2)]}}, \quad \quad \quad |J_3|
 \to  \infty ,
\end{equation}
and meet each other at
\begin{equation}
\label{Tmin}
T_c^{min} = \frac{2}{\ln{[q/(q-2)]}}, \quad \quad \quad J_3 = 0.
\end{equation}
At $J_3 = 0$ the system has a second critical point
\begin{equation}
\label{Tsec}
T_c^{sec} = \frac{2n}{\ln{[q/(q-2)]}}. \quad \quad \quad J_3 = 0.
\end{equation}

The phase diagrams ($T_c$ versus $\delta$) of the Ising model on the
two-layer Bethe lattice for different values $n = 1, 0.75, 0.5, 0.25,
0.1$ and for different values of coordination number $q =3, 4, 6, \infty$
are shown in Figs. 2(a), 2(b), and 2(c).
A few comments are in order. For $J_3 = 0$ we recover two critical
temperatures of two single-layer Bethe lattices with different intralayer
ferromagnetic coupling constants ($J_1$ and $J_2$).
In the opposite limit of $|J_3| \to \infty$, the critical temperature goes
asymptotically to a value given by Eq. (\ref{Tmax}) with the effective
intralayer coupling constant $J_1(1+n)$,
since the interlayer pairs become rigidly correlated.

\begin{center}
V. WEAK AND STRONG INTERLAYER COUPLING REGIMES
\end{center}

The spin-1/2 Ising model on a two-layer square lattice is
exactly soluble only in the cases $J_3 = 0$ and $|J_3| \to \infty$,
where it is related to the one layer square Ising model.
 When $J_3 = 0$, the Hamiltonian given by Eq. (\ref{R1}) describes two
uncoupled Ising model or, equivalently, two free fermionic fields. In strong
vertical interaction limits $|J_3| \to \infty$, each pair of spins
coupled across the layers will act as a single spin, and the Hamiltonian
given by Eq. (\ref{R1}) describes an one-layer Ising model with ($J_1 + J_2$)
as the coupling constant.

In the weak interlayer coupling regime ($J_3 \to 0$) the shift exponent 
$\psi$ can be defined by
\begin{equation}
\label{shift}
\Delta T_c \equiv \frac{T_c(J_3)-T_c(0)}{T_c(0)} \sim
 |J_3/J_1|^{\frac{1}{\psi}},
\end{equation}
where $T_c(J_3)$ is the critical temperature when the system has interlayer
coupling constant $J_3$.

In this section we calculate exactly the shift exponent for the
Ising model on a two-layer Bethe lattice.
In the weak coupling regime we obtain
\begin{equation}
\Delta T_c =b_1(q) \frac{|J_3|}{J_1} \quad \quad
\mbox{for} \quad \quad J_1 = J_2 \quad (n = 1)
\label{shift1}
\end{equation}
and
\begin{equation}
\Delta T_c = b_2(q,n) \left(\frac{J_3}{J_1}\right)^2 \quad \quad
\mbox{for} \quad \quad J_1 > J_2 \quad (n < 1)
\label{shift2}
\end{equation}
where
$$
b_1(q) = \frac{1}{q-2} \quad \mbox{and} \quad
b_2(q,n) = \frac{\ln{a}}{8(q-1)}
\left(\frac{1+a^{n-1}}{1 - a^{n-1}}\right) (a^{2n}-1),
$$
with  $a=q/(q-2)$.

Thus we find that the shift exponent $\psi$ for the system with
$J_1 = J_2$ is equal to $1$, which
coincides with theoretically predicted results $\psi = \gamma = 1$ 
for the single-layer Bethe lattice. For the system with
$J_1 \ne J_2$, we find that $\psi = 0.5$, which also exactly coincides with
the value predicted by the scaling theory \cite{ht97}.

In the strong coupling regime we have:

\begin{equation}
\frac{T_c(J_3)}{T_c^{max}} = 1 - K \exp{\left(-\frac{2|J_3/J_1|}
{T_c^{max}}\right)}
 \quad \quad
\mbox{for} \quad \quad J_1 \ge J_2 \quad (n \le 1)
\label{shiftnew}
\end{equation}
where 
$$
K = \frac{2(q-1)}{\ln{[q/(q-2)]}}(1-b^2)b^{q-2}
$$
with 
$$
b = \frac{q-2}{2(q-1)}\left[\left(\frac{q}{q-2}\right)^{\frac{1}{n+1}}+
\left(\frac{q}{q-2}\right)^{\frac{n}{n+1}}\right].
$$
It is easy to see from Eqs.(\ref{shift1}), (\ref{shift2}) and 
(\ref{shiftnew}) that the behavior of the strong coupling expansions is very 
different from the behavior in the weak coupling regime. It seems that we are 
the first group to obtain Eq. (\ref{shiftnew}) for the two-layer system 
with different intralayer coupling constants ($J_1 \ne J_2$). 
It should be noted that for the case $J_1 = J_2$, equations similar to
Eq. (\ref{shiftnew}) had been obtained by approximate methods
\cite{lsu93,acpv92}.

\begin{center}
VI. SUMMARY AND DISCUSSION
\end{center}
In the present paper we have investigated an Ising model on a bilayer
Bethe lattice with intralayer coupling constants $J_1$ and $J_2$ for
the first and the second layers, respectively, and interlayer coupling
constant $J_3$ between the two layers. We first analyze phase diagrams of
ground states, then using an iteration technique to obtain exact 
expressions for order parameters
and the free energy of the bilayer Ising model (Eqs. (18)-(21)).
We then obtain exact phase diagrams of Eqs. (30) and (31) and
analyze these equations in the weak and strong interlayer coupling
regimes, see Eqs. (36)-(38). The shift exponents $\psi$ in Eqs. (36)
and (37) are the first exact result to  support the scaling theory
for $\psi$, which states that $\psi$ is equal to the exponent
of magnetic susceptibility for $J_1 = J_2$ and is equal to 0.5
for $J_1 \ne J_2$ \cite{abe70,suzuki71,ht97}. It seems that
Eq. (38) is a new result.

In Sec. II, we present very rich phase diagrams for ground states.
However, in Sec. IV we consider only phase diagrams for $J_1 \ge J_2 > 0$.
It is of interest to study the evolution of phase  diagrams in Sec. II
as the temperature increases from 0 to high temperatures.
However, the analysis of such general phase diagrams is quite
complicated.

The dependence of various quantities on the film thickness is a
topic of current interest. In principle, we can extend our calculations 
from two layers to $n$ layers. For such a $n$-layer system, we can
introduce $n$ external magnetic fields $h_1$, $h_2$, $\dots$, $h_n$
($h_i$ for the $i-$th layer with $1 \le i \le n$), 
$C^n_2 (=n(n-1)/2!)$ interlayer coupling constants for
two-layer coupling, $C^n_3$ interlayer coupling constants for 
three-layer coupling, $\dots$, and $C^n_n (=1)$ coupling constant 
for $n$-layer coupling. Therefore, the total number of such
{\it field} coupling parameters
are $C^n_1+C^n_2+C^n_3+\dots+C^n_n=2^n-1$. Equations (15)-(17) for
two-layer systems for three {\it field} coupling parameters $(h_1,h_2,J_3)$
can be extended to $2^n-1$ equations for $2^n-1$ {\it field} 
coupling parameters. It is very difficult to find analytic or 
numerical solutions of these equations for $n > 2$.
However, we can simplify the problem by reducing the number of
independent {\it field} coupling parameters, i.e. setting
$h_i=h$ for $1 \le i \le n$ and keeping only a two-layer 
interlayer coupling parameter for two nearest-neighbor layers.
We are working in this direction.

\begin{center}
ACKNOWLEDGMENTS
\end{center}
We would like to thank J. G. Dushoff for a critical reading of the paper.
This work was supported by the National Science Council of the Republic of
China (Taiwan) under grant numbers NSC 87-2112-M-001-046. One of us
(N. Sh. I.) thanks the ISF Supplementary Grant - SDU000 and INTAS - 96 - 690
for partial financial support.

\figure{The ground-state phase diagram of the two-layer Ising
model for (a) $J_1 > 0$ and (b) $J_1 < 0$. }

\figure{Phase diagram on $(J_3/qJ_1,T/q)$ plane for an Ising model
on a two - layer  Bethe lattice with intralayer coupling constants
$J_1$ and $J_2$ for the  first and the second layer, respectively,
and interlayer coupling constant $J_3$ between two layers; here
$q$ is the coordination number for one-layer Bethe lattice and
$T=1/J_1 > 0$. (a) $J_1 = J_2$ and $q=3$.
A first-order phase boundary (dashed line) separates two ordered phases
designated by (F) and (C). The solid line
denotes the second order phase transition line, which separates paramagnetic
phase (P) from two ordered phases (F) and (C).
Notice that $T_c$ is the critical temperature of the Ising model
on one-layer Bethe lattice. (b) $J_1 = J_2$ and $q$=3, 4, 6, and $\infty$.
(c) $q=3$ and $n=J_2/J_1$=1, 0.75, 0.5, 0.25, and 0.1, which are denoted
on curves by 1, 2, 3, 4, and 5, respectively. $T_2$, $T_3$, $T_4$,
and $T_5$ are second  critical point of Eq.(34) for curves labeled by
 2, 3, 4, and 5, respectively.}


\begin{references}
\bibitem{J123} For the Ising model on a bilayer square
 (or other planar) lattice, we still use $J_1$ and $J_2$ to
 represent intralayer coupling constants for the first and the second
 layers, respectively, and use $J_3$ to represent the interlayer
 coupling constant.
\end{references}
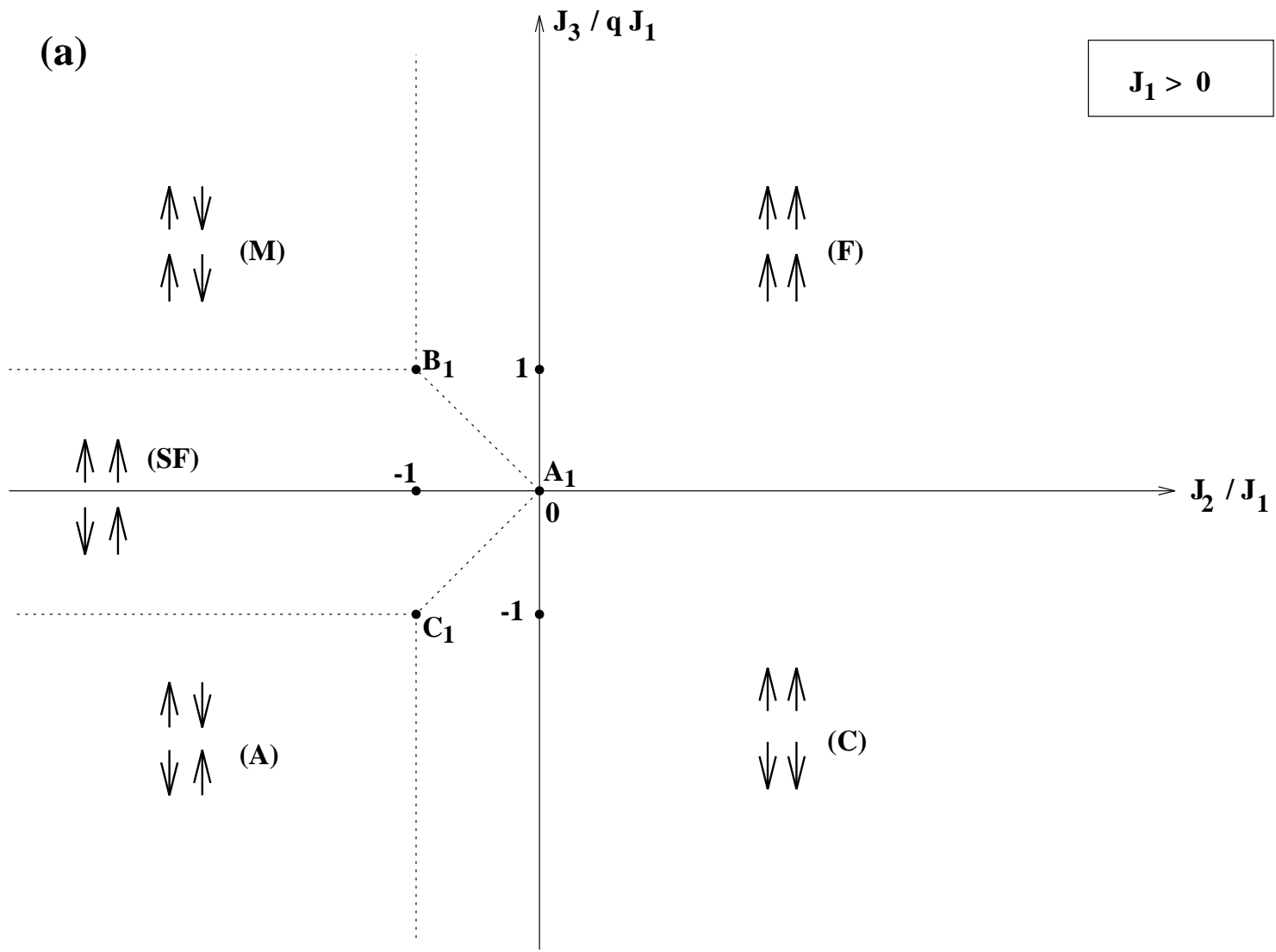
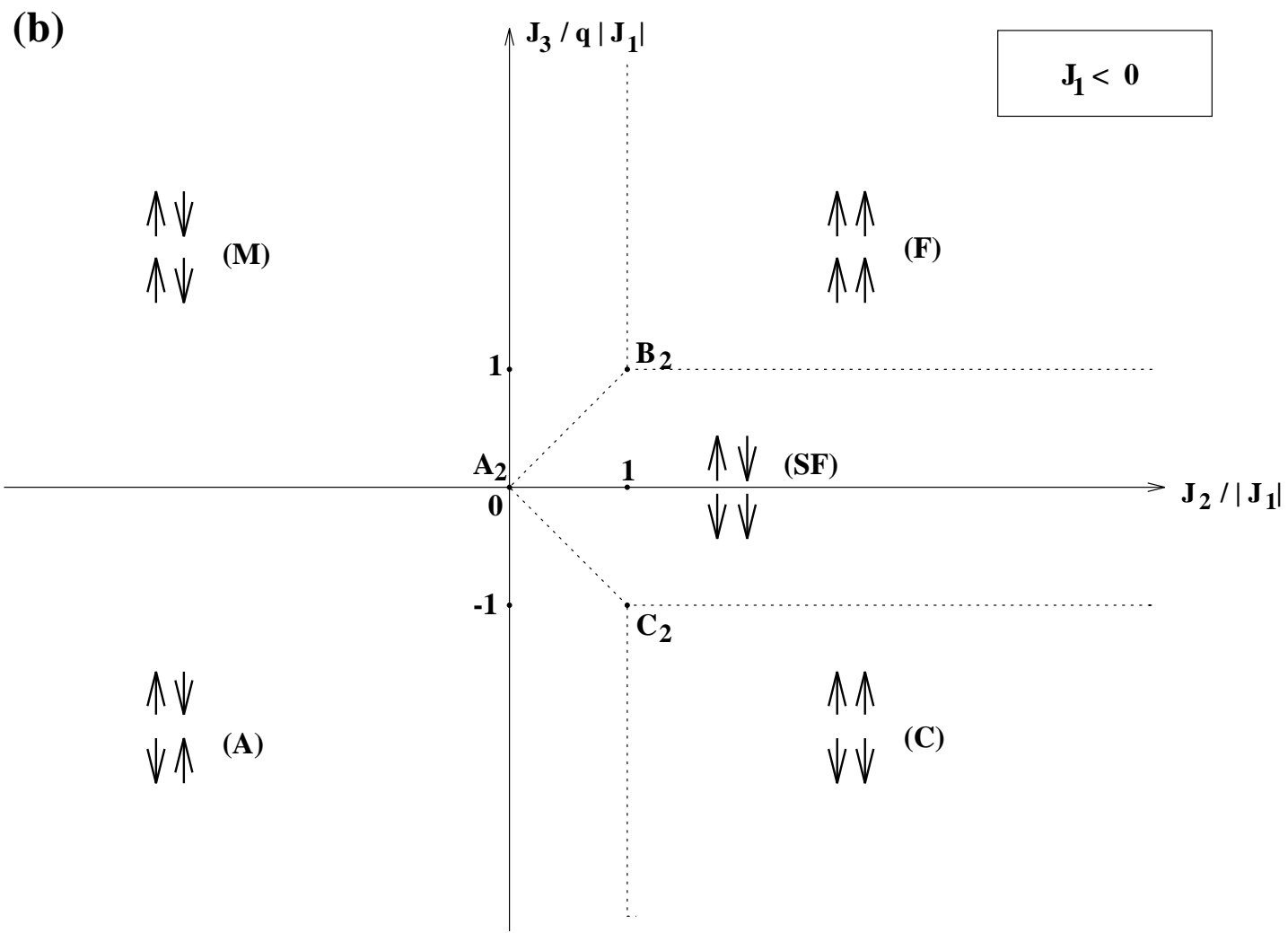
\end{document}